\documentstyle[preprint,prl,aps]{revtex}
\begin{document}
\draft
%
%
\title{On the EPR Phenomenon}
%
%
\author{Toshifumi Sakaguchi}
%
%
\address{
ASCII Corporation \\
Yoyogi 4-33-10, Shibuya-Ku, Tokyo, Japan
}
\date{\today}
\maketitle
%
%
\begin{abstract}
The origin of the nonlocal nature of quantum mechanics is investigated
in the context of Everett's formulation of quantum mechanics.  EPR
phenomenon can fully be explained without introducing any kind of
decoherence.
\end{abstract}
%
%
\narrowtext
%
%
 From the time when Einstein, Podolsky and Rosen argued~\cite{rf:epr}
the incompleteness of quantum mechanics by showing so-called EPR
thought experiment, there has been heated discussions~\cite{rf:book} on
the nonlocal nature of quantum mechanics.  Among them, Bell
proposed~\cite{rf:bell} an inequality which has to be satisfied if the
nature has a locality, and many experimental tests have subsequently
been
performed.~\cite{rf:exp1,rf:exp2,rf:exp3,rf:exp4,rf:exp5,rf:exp6,rf:exp7}

According to the experiments, the inequality is really violated, and so
we have to conclude that the nature possesses a {\it nonlocality\/}.
But how can this be possible by {\it local\/} interaction?  One of the
answers could be ``that is quantum mechanics''.  But if one takes
Everett's formulation of quantum mechanics, it can be understood from a
more fundamental level.  This paper is aimed at this point.  I will not
introduce any new interpretation to explain the phenomenon.  The only
assumption the formulation is based upon is that the physical law which
governs microscopic phenomena is also applicable
to macroscopic human beings.  Those who are not familiar with
Everett's formulation of quantum mechanics should read his original
paper.~\cite{rf:everett}

We use the same notation as the previous paper~\cite{rf:me} for
describing observer state:  An observer state in which an
observed value $\alpha^{i}$, one of the eigenvalues of an observable
$\alpha$, is recorded, is written as
$| [\alpha^{i}] \rangle$.  When an object system is in a superposition
$| \psi \rangle = \sum_{i} C_{i} | \alpha^{i} \rangle$,
the measurement process can be described as follows:
\begin{eqnarray}
U(t) |\psi \rangle|[\:] \rangle = \sum_{i} C_{i}
|\alpha^{i} \rangle|[\alpha^{i}] \rangle,
\label{eq:1}
\end{eqnarray}
where $U(t)$ is a time evolution operator obtained from a Hamiltonian
which includes the interaction between the object and observer
systems.  It means that when the observer observes the object system in
the state $|\psi\rangle$, the observer state $|[\:]\rangle$ branches
into a number of different observer states $|[\alpha^{i}] \rangle$,
each of which describes independent observer.  This is a direct
consequence of {\it the linearity of the time evolution operator
$U(t)$\/} and of {\it the relation which should be satisfied if the
object system is prepared in an {\it eigenstate} $|\alpha^{i} \rangle$
and the observer system is prepared to observe
the observable $\alpha$\/}:
\begin{eqnarray}
U(t) |\alpha^{i} \rangle|[\:] \rangle = |\alpha^{i} \rangle|[\alpha^{i}]
\rangle.
\label{eq:2}
\end{eqnarray}
When the observer system comes into interaction with the object system
for a second time, we can apply the time evolution operator to the
branched states again and we get a superposition of states
$|\alpha^{i} \rangle | [ \alpha^{i} \alpha^{i} ] \rangle$,
where the same value with
the one observed first is recorded in each observer's memory.
Therefore, it will {\it appear\/} to each observer, who is described by
each observer state in the superposition, that the {\it object\/}
system in the state $| \psi \rangle$ has {\it collapsed\/} into one of
the eigenstates of the observable $\alpha$.  The
collapse-of-wavefunction {\it phenomenon\/} of the {\it object\/}
system described above is nothing but the branching process of the
{\it observer\/} state according to Schr\"{o}dinger's equation, which
can never happen in classical physics.  It plays an essential role in
the EPR phenomenon.  That is, the nonlocal nature comes from the fact
that a state of a pair of observers {\it branches\/} into a number of
states of mutually correlated pairs by the {\it local interaction\/}
between the object and observer systems. This is shown below in detail.

We consider a two particle system in a state:
\begin{eqnarray}
\sum_{ij} C_{ij} | \alpha^{i} \rangle_{1} | \alpha^{j} \rangle_{2}.
\label{eq:3}
\end{eqnarray}
The two particles can be separated at any distance from each other.
When the observer $1$ observes the observable $\alpha$ of the
particle $1$ prepared in the state $| \alpha^{i} \rangle_{1}$,
the measurement process will be described as:
\begin{eqnarray}
U_{1} | \alpha^{i} \rangle_{1} | \: \rangle_{2}
| [\:] \rangle_{1} | [\:] \rangle_{2}
=  | \alpha^{i} \rangle_{1} | [\alpha^{i}] \rangle_{1}
| \: \rangle_{2} | [\:] \rangle_{2}.
\label{eq:4}
\end{eqnarray}
It should be noted that the measurement process is local and it does not
affect the state of the system $2$.  When the observer $2$ observes an
observable $\beta$, which does not commute with
$\alpha$, of the particle $2$ prepared in the state
$|\alpha^{j} \rangle_{2}$, the measurement process will be described as:
\begin{eqnarray}
&U_{2}& | \: \rangle_{1} | \alpha^{j} \rangle_{2}
| [\:] \rangle_{1} | [\:] \rangle_{2} \nonumber \\
&&= | \: \rangle_{1} | [\:] \rangle_{1}
\sum_{j^{\prime}}
{}_{2} \langle \beta^{j^{\prime}} | \alpha^{j} \rangle_{2}
|\beta^{j^{\prime}} \rangle_{2}
|[\beta^{j^{\prime}}] \rangle_{2}.
\label{eq:5}
\end{eqnarray}
That is, the state of the system $1$ remains unaffected, while
the state of the observer $2$ branches into the states
$|[\beta^{j^{\prime}}] \rangle_{2}$.
The state vector $|\Psi \rangle$ of the whole system after
the two observations performed therefore becomes
\begin{eqnarray}
| \Psi \rangle &=& U_{1} U_{2} \sum_{ij} C_{ij}
| \alpha^{i} \rangle_{1} | \alpha^{j} \rangle_{2}
| [\:] \rangle_{1} | [\:] \rangle_{2}
\nonumber \\
&=& \sum_{i j^{\prime}}
|\alpha^{i} \rangle_{1}|[\alpha^{i}] \rangle_{1}
\{ \sum_{j} C_{ij}
{}_{2}\langle \beta^{j^{\prime}} | \alpha^{j} \rangle_{2} \}
|\beta^{j^{\prime}} \rangle_{2} |[\beta^{j^{\prime}}] \rangle_{2}
\nonumber \\
&=& \sum_{i j^{\prime}} K_{i j^{\prime}}
| \alpha^{i} \beta^{j^{\prime}} \rangle
| [(\alpha^{i} \beta^{j^{\prime}})] \rangle,
\label{eq:6}
\end{eqnarray}
where
$K_{i j^{\prime}} \equiv \sum_{j} C_{ij} {}_{2}\langle
\beta^{j^{\prime}} | \alpha^{j} \rangle_{2}$,
$|\alpha^{i} \beta^{j^{\prime}} \rangle
\equiv | \alpha^{i} \rangle_{1} | \beta^{j^{\prime}} \rangle_{2}$ and
$| [(\alpha^{i} \beta^{j^{\prime}})] \rangle
\equiv | [\alpha^{i}] \rangle_{1} | [\beta^{j^{\prime}}] \rangle_{2}$.

Preparing $N$ identical pairs of particles in the same state
(\ref{eq:3}) and performing the above observation sequentially,
we will get a superposition of branched states of the form
\begin{eqnarray}
K_{i i^{\prime}} K_{j j^{\prime}}
&& \ldots K_{k k^{\prime}}
| \alpha^{i} \beta^{i^{\prime}} \rangle
| \alpha^{j} \beta^{j^{\prime}} \rangle
\ldots
| \alpha^{k} \beta^{k^{\prime}} \rangle
\nonumber \\
&& \otimes
| [(\alpha^{i} \beta^{i^{\prime}})
(\alpha^{j} \beta^{j^{\prime}})
\ldots
(\alpha^{k} \beta^{k^{\prime}}) ] \rangle,
\label{eq:7}
\end{eqnarray}
where
\begin{eqnarray}
| [(\alpha^{i} \beta^{i^{\prime}})
(\alpha^{j} \beta^{j^{\prime}}) &&\ldots
(\alpha^{k} \beta^{k^{\prime}}) ] \rangle \nonumber \\
&&\equiv | [\alpha^{i} \alpha^{j} \ldots \alpha^{k}] \rangle
| [\beta^{i^{\prime}} \beta^{j^{\prime}} \ldots
\beta^{k^{\prime}}] \rangle.
\label{eq:8}
\end{eqnarray}
In the limit $N \rightarrow \infty$, we can show~\cite{rf:me} that
the pair $(\alpha^{p} \beta^{p^{\prime}})$ in
$[ \ldots ]$ appears $| K_{p p^{\prime}} |^{2} N$
times in almost all branched states in the superposition.
Therefore, each pair of observers in the superposition
will conclude that the pair $(\alpha^{p} \beta^{p^{\prime}})$
can be obtained with probability
\begin{eqnarray}
P(\alpha^{p}, \beta^{p^{\prime}}) &=& | K_{p p^{\prime}} |^{2}
\nonumber \\
&=& \sum_{qr} C^{\ast}_{pq} C_{pr}
\langle \alpha^{q} | \beta^{p^{\prime}} \rangle
\langle \beta^{p^{\prime}} | \alpha^{r} \rangle.
\label{eq:9}
\end{eqnarray}
If the state (\ref{eq:3}) is the spin singlet state of spin-1/2
particles;
\begin{eqnarray}
\bigl( C_{ij} \bigr) =
\frac{1}{\sqrt{2}}
\left(
\begin{array}{cc}
 0 & 1 \\
-1 & 0
\end{array}
\right),
\label{eq:10}
\end{eqnarray}
and
\begin{eqnarray}
\bigl( \langle S^{i}_{z} | S^{j}_{z^{\prime}} \rangle \bigr) =
\left(
\begin{array}{cc}
\cos \theta / 2 & - i \sin \theta / 2 \\
- i \sin \theta / 2 & \cos \theta / 2
\end{array}
\right),
\label{eq:11}
\end{eqnarray}
where $\theta$ is the angle between the $z$-axis and $z^{\prime}$-axis,
the probability becomes
\begin{eqnarray}
\bigl( P(S^{i}_{z}, S^{j}_{z^{\prime}}) \bigr) =
\frac{1}{2}
\left(
\begin{array}{cc}
\sin^{2} \theta / 2 & \cos^{2} \theta / 2 \\
\cos^{2} \theta / 2 & \sin^{2} \theta / 2
\end{array}
\right).
\label{eq:12}
\end{eqnarray}

The branching process caused by the local interaction between the
object and observer systems is local and does not affect the
observation performed by the other observer at a remote site as shown
in Eqs.~(\ref{eq:4}) and (\ref{eq:5}). This is why superluminal
communication is impossible.  The probability derived above, which is
shown to violate Bell's inequality, can be recognized only when the two
observers meet each other to compare the data in their world.
%
%

%
%
%
\end{document}